\documentclass[]{spie}  

\usepackage{amsmath,amsfonts,amssymb}
\usepackage{graphicx}
\usepackage[colorlinks=true, allcolors=blue]{hyperref}
\usepackage{booktabs}

\def\micron{$\mu$m}

\def\copol{Co-pol}
\def\xpol{X-pol}
\def\leakage{Leakage}
\def\reflection{Reflection}

\title{Tolerance Analysis of Octave Bandwidth Millimeter-Wave Planar Orthomode Transducer}

\author[a]{Johannes~Hubmayr}
\author[a]{Jason~E.~Austermann}
\author[a]{James~A.~Beall}
\author[a]{Jake~A.~Connors}
\author[a]{Shannon~M.~Duff}
\author[b]{Jeffrey~J.~McMahon}
\affil[a]{NIST Quantum Sensors Group, 325 Broadway Ave, Boulder, CO, 80305, USA}
\affil[b]{Department of Astronomy and Astrophysics, University of Chicago, 5640 S. Ellis Ave, Chicago, IL 60637, USA}

\authorinfo{Corresponding author: J. Hubmayr (hubmayr@nist.gov)}

\begin{document} 
\maketitle

\begin{abstract}
Planar Orthomode Transducers (OMTs) are commonly used for polarization measurements at millimeter wavelengths.  
We present an optical coupling study of an octave bandwidth planar OMT in circular waveguide based on 3D electromagnetic simulations.  
We quantify results through metrics such as co- and cross- polar coupling, reflection, and waveguide leakage as a function of the OMT construction geometry.  
We evaluate the tolerance of these metrics to the waveguide backshort distance, probe impedance, waveguide gap size, and waveguide-to-probe misalignment.    
Two probe geometries are studied: the `classic' shape used in several previous experiments, and a new `wineglass' geometry.  
The bandwidth ratio of both optimized OMTs is 2.0:1, defined where co-polar coupling exceeds 80\%.  
The average co-polar coupling, cross-polar coupling, reflection, and waveguide leakage of the classic probe is approximately 93\%, $<$~-50~dB, 5\% and 2\%, respectively and depends slightly on the exact frequency range.  
The wineglass probe co-polar coupling is $\sim$~2\% larger.  
Radial waveguide misalignment at the level of 4\% of the waveguide radius can result in up to a 10\% reduction in co-polar coupling and -20~dB cross-polar coupling in one polarization.   
These results may be used to guide the detector module designs of future Cosmic Microwave Background experiments and beyond.

\end{abstract}

\keywords{planar OMT, orthomode transducer, CMB}

\section{INTRODUCTION}
\label{sec:intro}  

%
A traditional orthomode transducer (OMT) is a three-dimensional waveguide component, which either separates or combines two orthogonally polarized microwave signals.   
In a planar OMT of the type described herein,  membrane supported polarization-sensitive probes typically made of a superconducting thin film are inserted into a waveguide cavity.  
Once on a planar structure, subsequent on-chip circuitry conditions the signal for power sensing.  
Planar OMTs of this type have found application in multiple Cosmic Microwave Background (CMB) polarization experiments
\cite{essinger2009atacama,stanchfield2016development,austermann2012sptpol,niemack2010actpol,hubmayr2016spider,essinger2014class} because they can separately measure orthogonal linear polarization states within one spatial pixel and can be fabricated into large arrays.  

The optimization and geometric tolerance analysis of 30\% fractional bandwidth planar OMTs within circular waveguide operating near 150~GHz has previously been studied \cite{mcmahon2009planar}.  
Octave bandwidth planar OMTs demonstrated at microwave frequencies \cite{grimes2007compact} have been frequency-scaled and integrated to form multichroic transition-edge-sensor (TES) polarimeters \cite{mcmahon2012multi}.  
Detectors of this type have been deployed in two generations of Atacama Cosmology Telescope (ACT) receivers, spanning the center frequency range 27--230~GHz\cite{thornton2016atacama, henderson2016advanced}.  
Multiple variations of this technology are either in production or in development for upcoming instruments \cite{ade2019simons,li2019probing,abazajian2019cmb,mazumi2020litebird}.   

Given their wide-spread adoption, here we present an optimization and tolerancing study for octave bandwidth planar OMTs based on electromagnetic simulations.  
As such this paper serves as an update to McMahon et al. 2009~\cite{mcmahon2009planar} for these broader bandwidth OMTs.  
Section~\ref{sec:model} describes the physical model, simulation parameters, and defines the optical coupling metrics.
Section~\ref{sec:results} presents the results of the optimized geometry and the variation of the metrics to the most relevant geometric parameters.  
We include band-averaged results within two bands since in most applications the bandwidth is partitioned to form a dichroic, dual-polarization sensitive pixel.  
In Sec.~\ref{sec:wine}, we present a new OMT probe geometry, which has been discussed in the context of CMB-S4, and compare the performance to the `classic' probe geometry of Grimes et al. 2007~\cite{grimes2007compact}.  
Discussion and conclusions are in Sec.~\ref{sec:discussion} and Sec.~\ref{sec:conclusion}, respectively.

\section{MODEL DESCRIPTION}
\label{sec:model}


\begin{figure} [t]
   \begin{center}
   \begin{tabular}{ccc} 
   \includegraphics[width=0.3\textwidth]{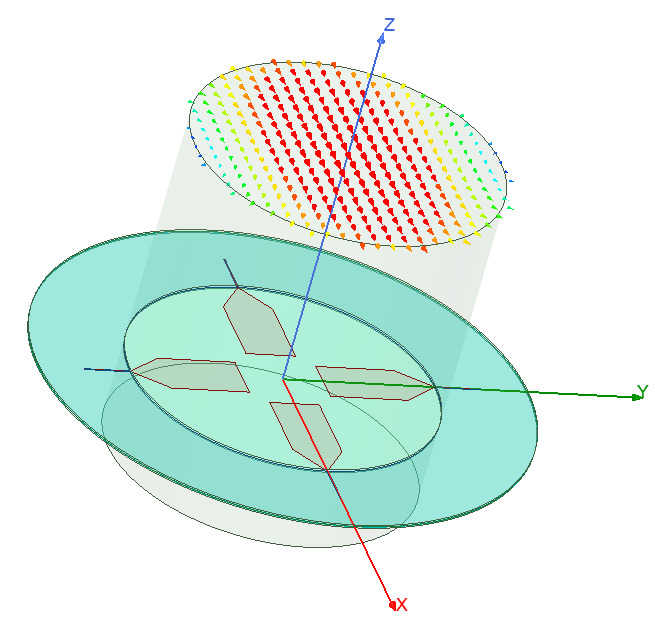} 
   \includegraphics[width=0.3\textwidth]{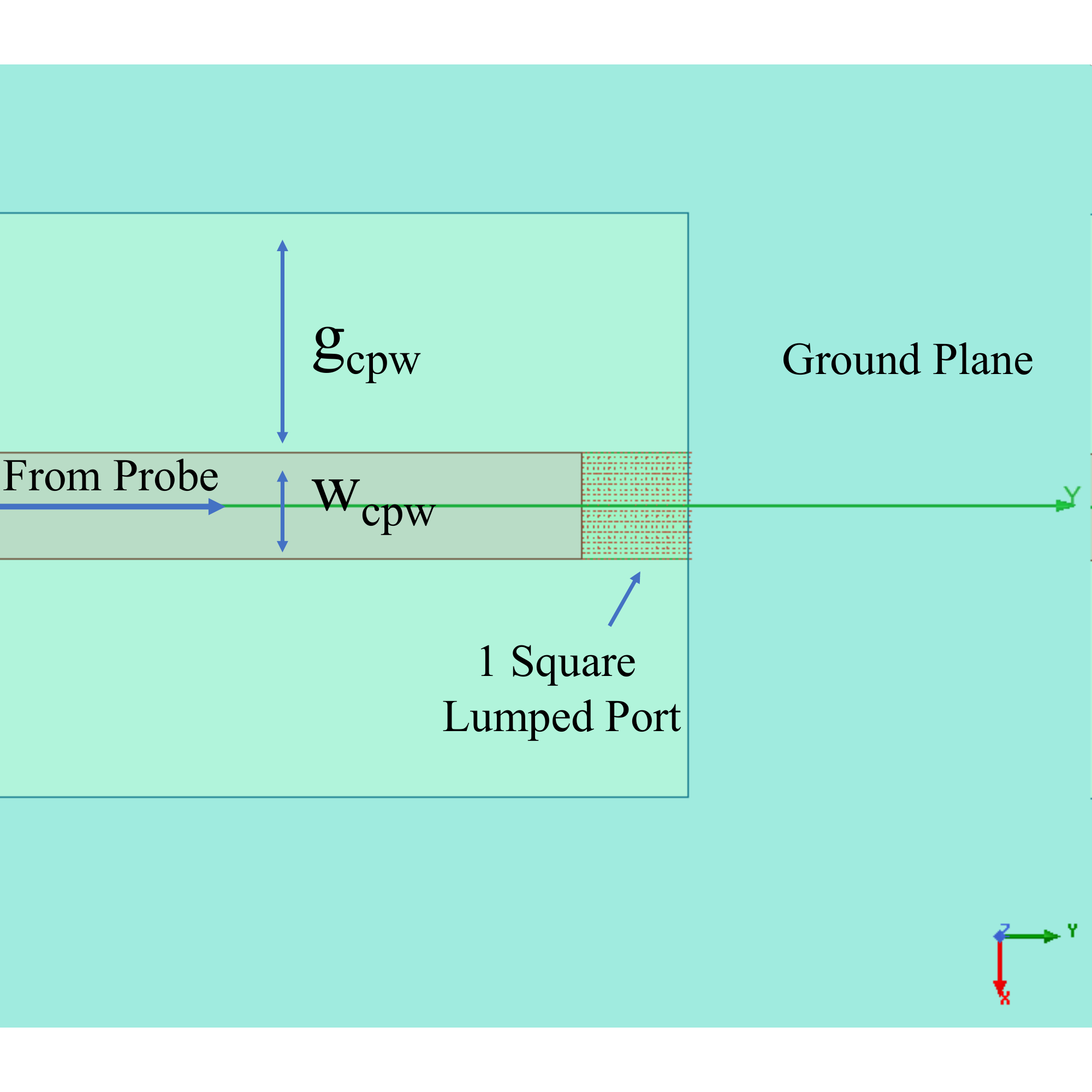} 
   \includegraphics[width=0.3\textwidth]{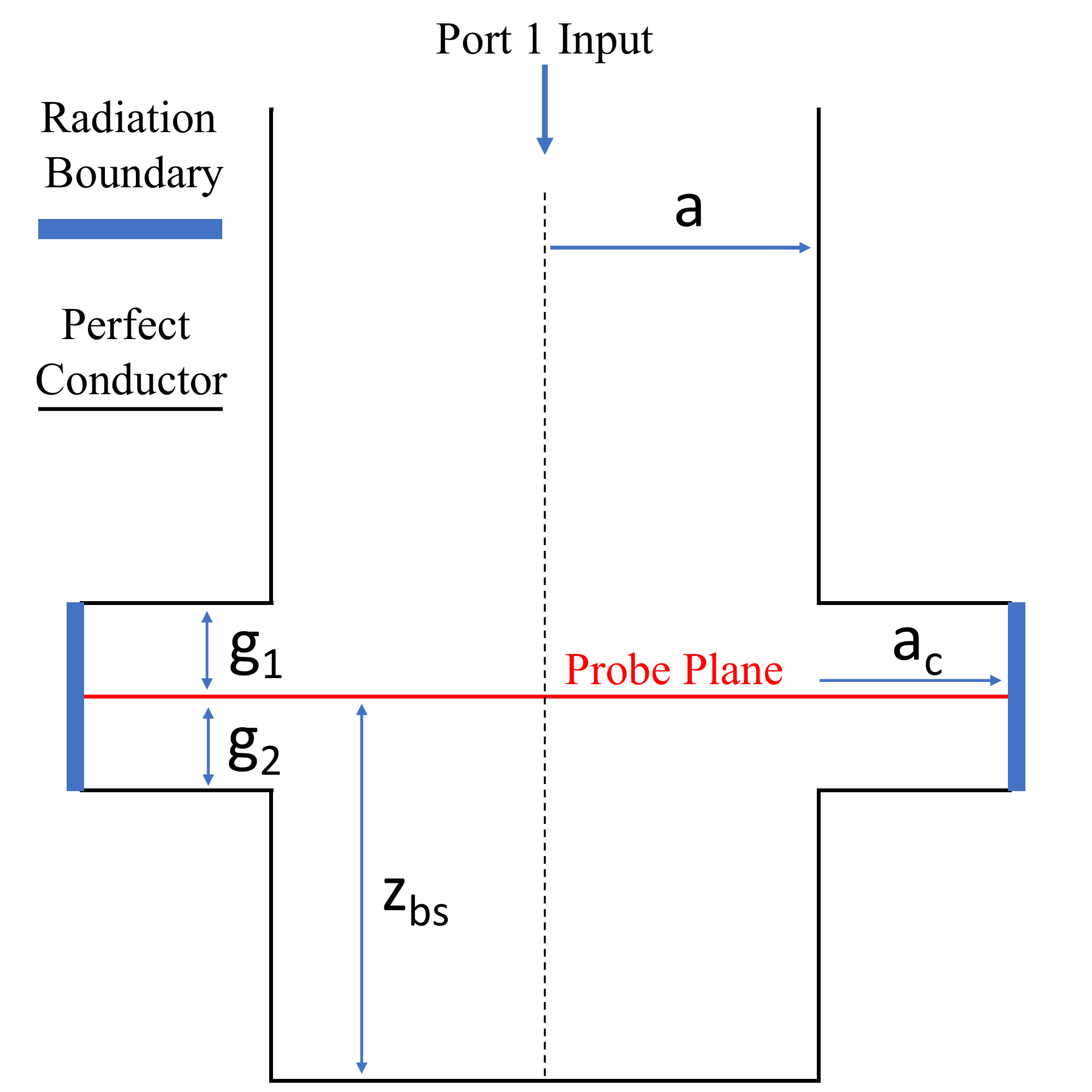} 
   \end{tabular}
   \end{center}
   \caption[example] 
   { \label{fig:model} 
   Left: Simulation volume with TE$_{11}$ mode superimposed over the input waveguide.
   Center: Expanded view of the lumped port used to terminate each of the four OMT probes. 
   Right: Cross-section of the simulation model with labeled components (not to scale).

}
   \end{figure} 

\begin{table}[b]
   \centering
   \begin{tabular}{@{} llll @{}} 
      \toprule
      \cmidrule(r){1-2} 
      Parameter    & Description & Value & Value/$a$ \\
      \midrule
      $a$ & waveguide radius & 1200~\micron & 1 \\
      $a_{\mathrm{c}}$ & ground-plane radial width & 725~\micron & 0.604 \\
      $z_{\mathrm{bs}}$ & backshort distance & 675~\micron & 0.5625 \\
      $g_1$ & gap to input (top) waveguide & 15~\micron & 0.0125 \\
      $g_2$ & gap to backshort (bottom) waveguide & 10~\micron & 0.0083   \\
      $h_\mathrm{p}$ & probe height & 390~\micron & 0.325 \\
      $\ell_\mathrm{p1}$ & probe length, rectangular section & 612~\micron & 0.51 \\
      $\ell_\mathrm{p2}$ & probe length, triangular section & 268.8~\micron & 0.224 \\
      \midrule
      $w_{\mathrm{cpw}}$ & CPW width & 2~\micron & \\
      $g_{\mathrm{cpw}}$ & CPW gap & 4.5~\micron & \\
      $t_{\mathrm{SiN_x}}$ & SiN$_\text{x}$ thickness & 2.0~\micron & \\
      $t_{\mathrm{SiO_2}}$ & SiO$_2$ thickness & 0.45~\micron & \\
      $t_{\mathrm{p}}$ & thickness of probe and ground plane & 0.2~\micron & \\  
      \midrule
      $\lambda$ & Nb London penetration depth & 0.085~\micron & \\
      $\epsilon_{\mathrm{r,SiN_x}}$ & SiN$_\text{x}$ relative dielectric constant & 6.8 & \\
      $\epsilon_{\mathrm{r,SiO_2}}$ & SiO$_2$ relative dielectric constant & 4.0 & \\
            \bottomrule
            
   \end{tabular}
   \vspace{0.1in}
   \caption{Planar OMT model parameters separated into three sections: geometrical parameters that scale with frequency proportional to the waveguide radius, geometrical parameters that do not scale with frequency, and material parameters. 
  }
   \label{tab:model_parameters}
\end{table}

\begin{figure} [t]
   \begin{center}
   \begin{tabular}{c} 
   \includegraphics[height=2.0in]{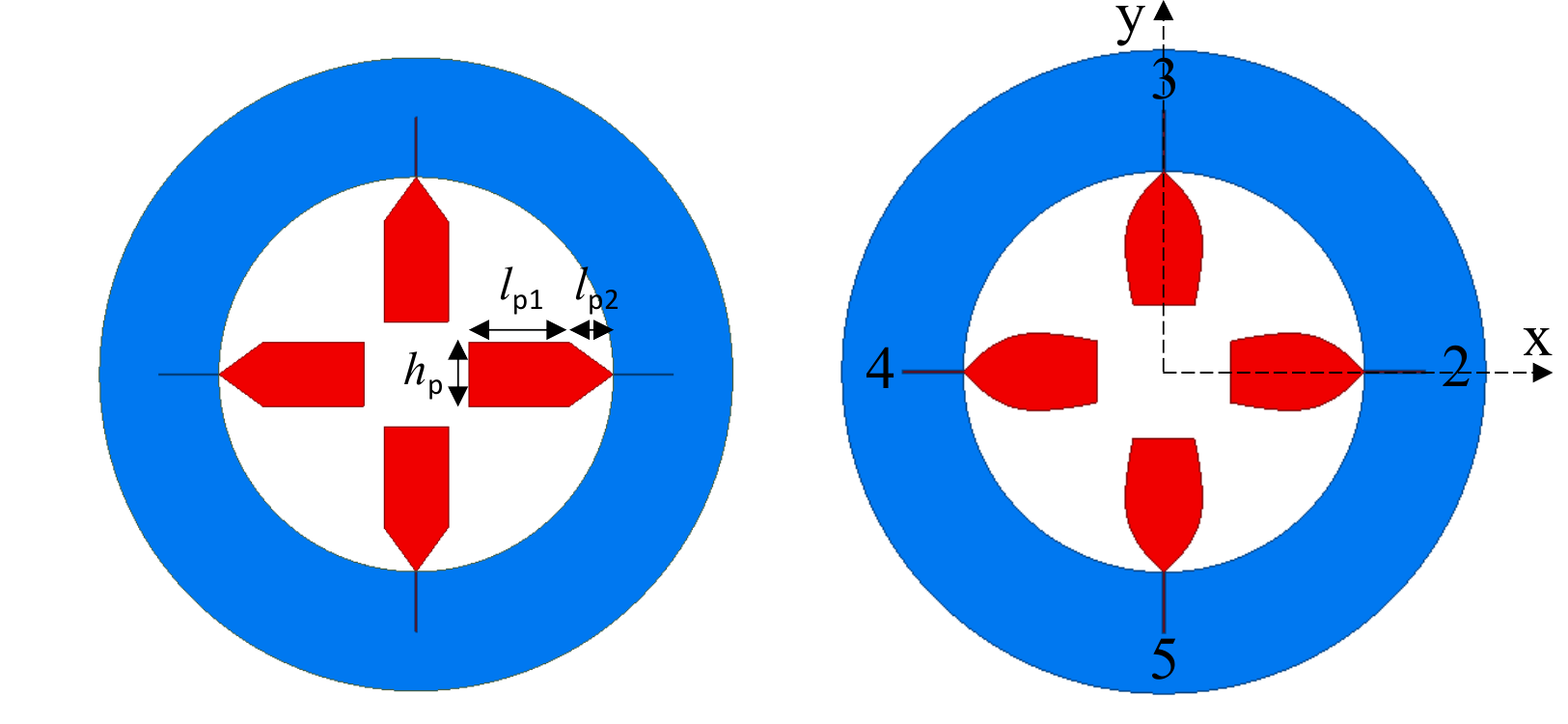}
   \end{tabular}
   \end{center}
   \caption[example] 
   { \label{fig:probe_geometry} 
   Geometry of `classic' (left) and `wineglass' (right) planar OMT probes.  
   Both red and blue areas are thin film Nb.  
   The port numbers are defined in the image on the right.
}
   \end{figure} 

We use ANSYS HFSS\footnote{The use of this software is not an endorsement by the US government.  We list by name to be comprehensive.} to optimize the planar OMT and determine the tolerance to geometric variations.
The physical model is shown in Fig.~\ref{fig:model} with labeled parameters that are defined and listed in Tab.~\ref{tab:model_parameters}.  
An input waveguide of radius $a$ feeds four planar OMT probes of geometry defined in Fig.~\ref{fig:probe_geometry} that lie a distance $z_{bs}$ away from a reflective backshort.  
The waveguide walls are modeled as perfect conductors.
Vertical gaps $g_1$ and $g_2$ from the probes to the top and bottom waveguide sections are necessary to insert the probes within the waveguide.  
For the summed height of the gaps and at a radial distance $a+a_c$, we implement a radiation boundary that allows the determination of the waveguide leakage, the power which escapes the waveguide.    
The physical realization of this geometry uses three separate silicon parts and is described in Ward et al. 2016\cite{ward2016mechanical}.

We model the OMT probes as 2D sheets with an impedance boundary $Z_{probe} = j \omega L$, where $$L = \mu_0 \frac{\lambda}{\tanh{(t_p/\lambda)}}.$$ 
$\omega$ is the angular frequency, $L$ is the kinetic inductance modeled with London penetration depth $\lambda~=~85~$nm (suitable for thin film Nb), and $t_\mathrm{p}$ is the thickness of the metal. 
Each probe output is a coplanar waveguide (CPW) of width $w_\mathrm{cpw}$  and gap $g_\mathrm{cpw}$, which is terminated in a lumped port of impedance $Z_\mathrm{term}=90~\Omega$ that matches the CPW impedance.  
As compared to waveports, the internal lumped ports provide a better representation of the true geometry and avoid possible low-level reflection artifacts at the expense of supporting only a single mode.      
The probes sit on a membrane comprised of a layer of SiN$_\text{x}$ and SiO$_2$ of thickness $t_\mathrm{SiN_x}$~=~2~$\mu$m and $t_\mathrm{SiO_2}$~=~0.45~$\mu$m, respectively and relative dielectric constant $\epsilon_\mathrm{r,SiN_x}$=6.8 and $\epsilon_\mathrm{r,SiO_2}$=4.0, respectively.  
These choices match the planar OMT fabrication process described in Duff 2016 \cite{duff2016advanced}.
We do not include loss in the dielectrics to aid our determination of waveguide leakage, as we define below.

We excite the input waveguide with eight modes, and align the first mode (TE$_{11}$) with the x-axis.  
For each parameter combination, we perform frequency sweeps that span a typical 90/150~GHz dichroic OMT.  
However, since results may be frequency-scaled as 1/$a$, we report both geometric parameters normalize to the waveguide radius $a$ and the frequency axis normalized by the waveguide cut-off frequency $f_\text{c}$.  
Thus, the frequency sweep is in the range 0.9~$<f/f_\mathrm{c}<~$2.38 with stepsize 0.054~$f/f_{c}$. 
The solution frequency is 2.05~$f/f_\mathrm{c}$.    

We largely use default HFSS parameters.  
The convergence criteria is standard, that successive passes change the S-parameters by $<0.02$.  
The maximum number of passes is set to 20.    
We use the direct solver and `first order' basis functions.
All of our simulations achieve the defined convergence criteria.  
Using a modest machine (4~GHz processor with 4 cores and 256~GB RAM), the average simulation time is 30s per frequency point.  

For simulations that require no waveguide misalignment, we make use of a quarter-sized model with appropriate E- and H- symmetry planes to substantially decrease the simulation time.  
Radial misalignment of the waveguide is incompatible with these boundary conditions, and thus a full model is used in these situations.
We find good agreement between the full and quarter model.  
For example the maximum difference in co-polarization is $<$~2\%.     

From the S-parameters of the eight modes of the input waveguide port (port 1) to the four lumped ports which terminate each OMT probe (ports 2-5 defined defined in Fig.~\ref{fig:probe_geometry}), we define the co- and cross- polar power coupling (\copol\ and \xpol), power reflection (\reflection), and waveguide leakage (\leakage) as follows:
\begin{eqnarray}
&& \mathrm{\copol} \overset{\mathrm{def}}{=}  |S(2,1:1)|^2 + |S(4,1:1) |^2 \label{eqn:copol} \\
&& \mathrm{\xpol} \overset{\mathrm{def}}{=} |S(2,1:2)|^2 + |S(4,1:2) |^2 \label{eqn:xpol} \\
&& \mathrm{\reflection} \overset{\mathrm{def}}{=}  |S(1:1,1:1)|^2 \label{eqn:s11} \\
&& \mathrm{\leakage} \overset{\mathrm{def}}{=} 1-\sum_{i=1}^8 |S(1:i,1:1) |^2-\sum_{j=2}^{5} |S(j,1:1)|^2 \label{eqn:leakage}
\end{eqnarray}
Here the convention $S(i:n,j:m)$ denotes the voltage at port $i$ in mode $n$ divided by the voltage produced at port $j$ in mode $m$.  
Since the lumped ports contain only a single mode, we drop the mode index for ports 2-5.  

Coupling to the TE$_{11}$ mode (mode $n$=1 in Eqns.~\ref{eqn:copol}-\ref{eqn:leakage}) is desirable, whereas coupling to higher order modes is not.  
This goal is represented by these definitions.  
Simply put, the design goal is to construct a planar OMT that maximizes \copol\ and minimizes \xpol, \reflection, and \leakage\ over the broadest bandwidth achievable.
As can be seen by these definitions, we sum the power in opposite probes incoherently, which is accomplished in real devices by ancillary on-chip circuitry.  
In the following section, we report these metrics as a function of model geometry.

\section{RESULTS}
\label{sec:results}

\begin{figure} [b]
   \begin{center}
   \begin{tabular}{c} 
   \includegraphics[width=\textwidth]{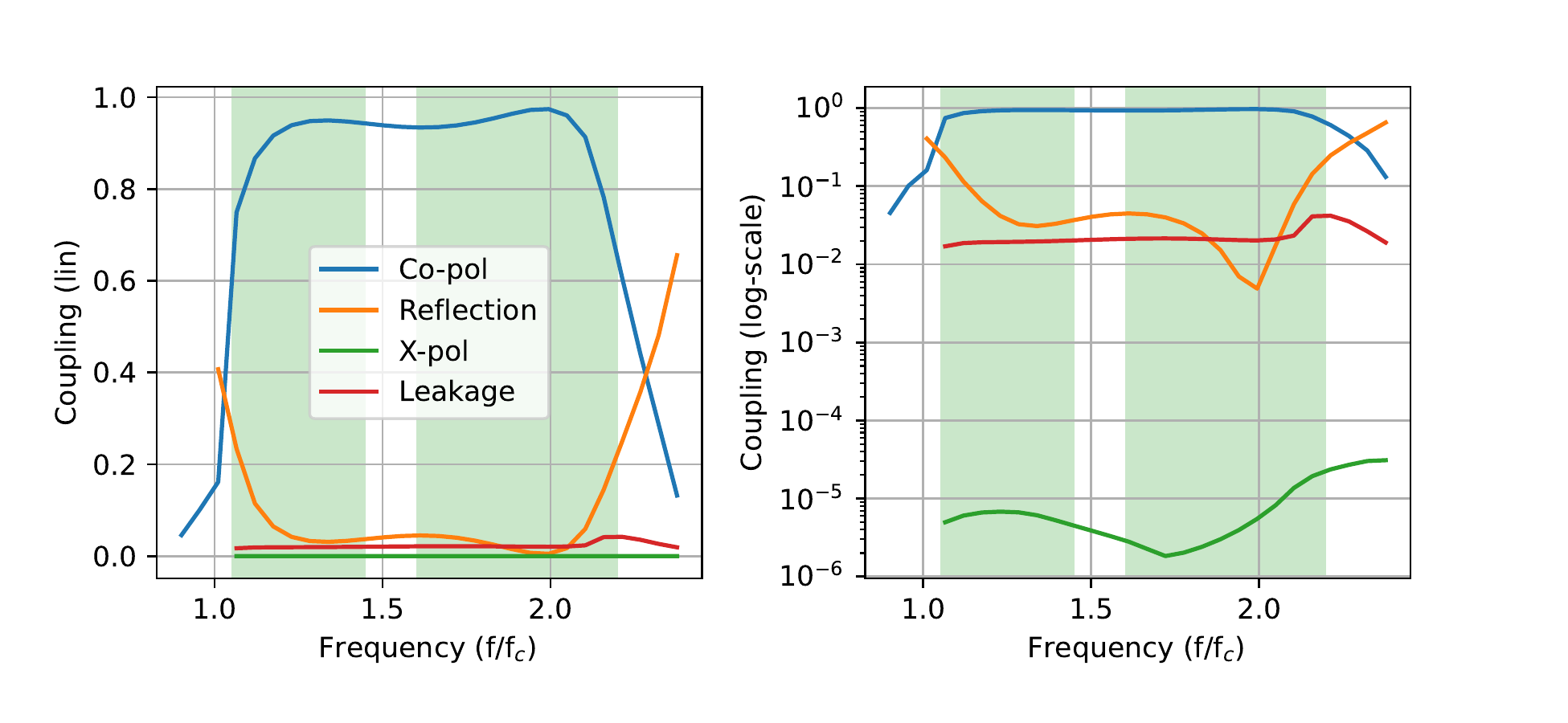}
   \end{tabular}
   \end{center}
   \caption[example] 
   { \label{fig:classic_coupling} 
Linear (left) and logarithmic (right) OMT coupling summary for the `classic' probes.  
Co- (cross-) polar coupling peaks at 96\% (-47~dB), and the leakage out of the waveguide is 
2\%.  
The full bandwith ratio of the OMT is 2.0:1, defined where co-pol~$>$~0.8.  
For guidance, the two shaded regions indicate fiducial bands in a dichroic OMT-coupled pixel, commonly used in ground-based CMB instruments.  
}
   \end{figure}

Figure~\ref{fig:classic_coupling} presents the coupling summary of the optimized planar OMT geometry.  
Since results scale with the waveguide radius $a$, and this has been explicitly confirmed in simulation, we report the frequency axis normalized by $f_c$.  
In this manner, results are applicable to any specific octave of bandwidth.
The bandwidth ratio, defined where \copol$~>0.8$, is 2.0:1.  
For guidance, the two shaded regions in Fig.~\ref{fig:classic_coupling} indicate fiducial bands (low band: 1.05$<f/f_\mathrm{c}<$1.45 and high band: 1.6$<f/f_\mathrm{c}<$2.2) commonly used in dichroic OMT-coupled pixels. 
Band averaged metrics defined in Eqns.~\ref{eqn:copol}-\ref{eqn:leakage} for the optimized geometry are reported in Tab.~\ref{tab:ave_coupling_values} along with these values for the 
`wineglass' probes discussed in Sec.~\ref{sec:wine}. 

\subsection{Higher-Order Mode Coupling}
\label{sec:modes}

\begin{figure} [t]
   \begin{center}
   \begin{tabular}{c} 
   \includegraphics[width=\textwidth]{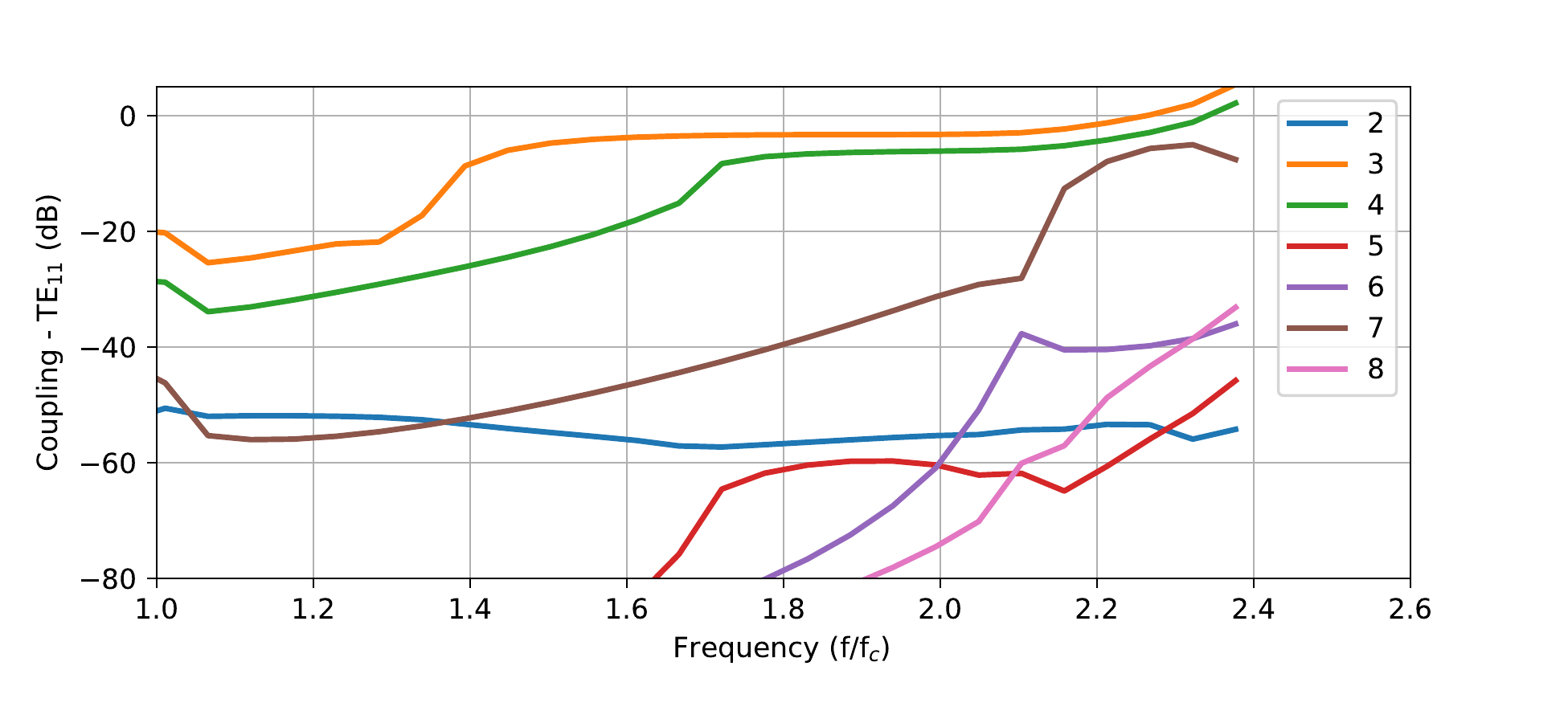}
   \end{tabular}
   \end{center}
   \caption[example] 
   { \label{fig:mode} 
Coupling of higher order modes to the `classic'-probe OMT relative to the TE$_{11}$ mode.  
The TM$_{01}$ (mode 3) and TE$_{21}$ (mode 4) are the only parasitic modes that couple stronger than -20~dB across the useable bandwidth.  }
\end{figure} 

\begin{figure} [b]
   \begin{center}
   \begin{tabular}{c} 
   \includegraphics[width=0.9\textwidth]{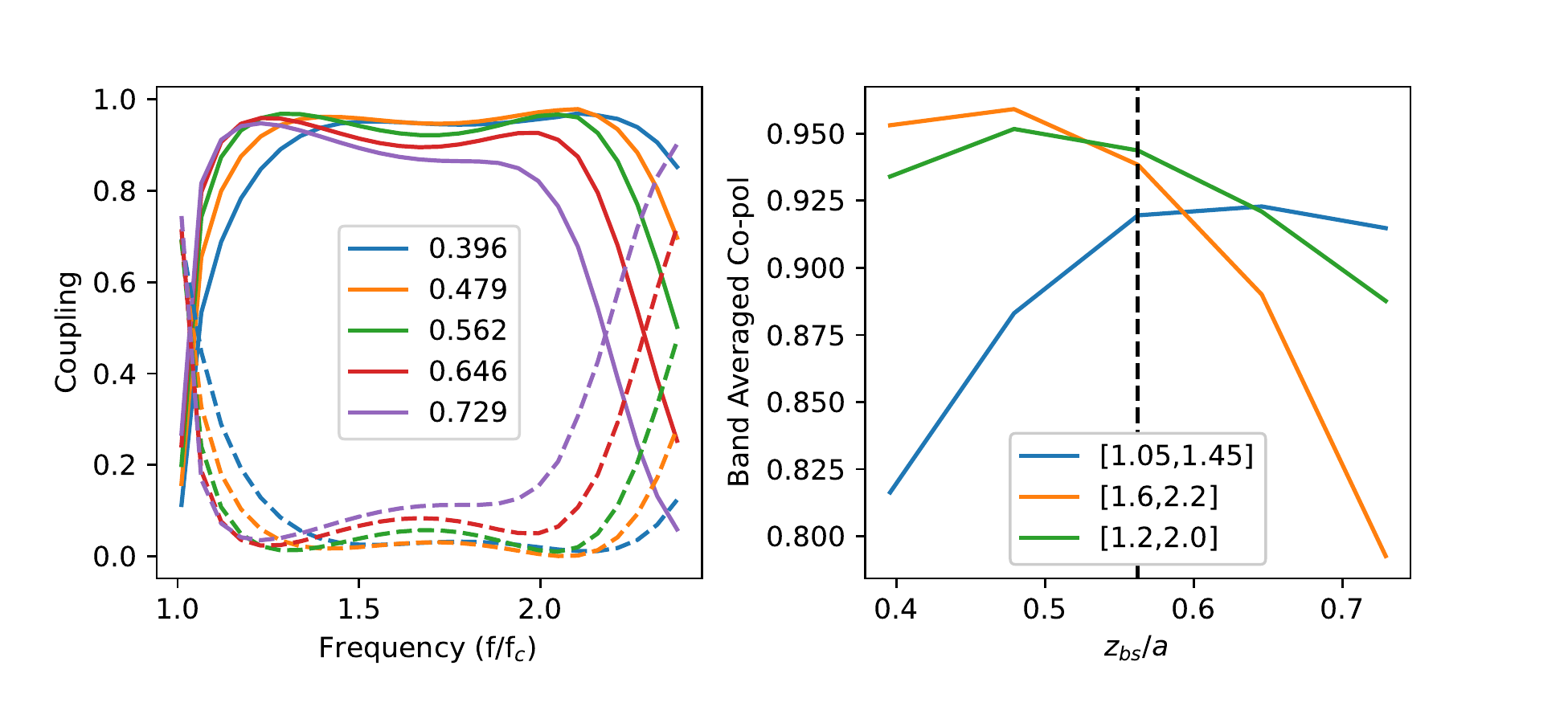}
   \end{tabular}
   \end{center}
   \caption[example] 
   { \label{fig:backshort} 
Effect of distance to reflective backshort $z_\mathrm{bs}$.  Left: \copol\ (solid) and \reflection\ (dashed) as a function of frequency for several 
backshort distances listed in the legend as a fraction of the waveguide radius $a$.  
Right: band-averaged co-polar coupling as a function of normalized backshort distance 
for three different frequency ranges listed in the legend.  
The vertical dashed-line indicates the fiducial value.
A 25\% change in $z_\mathrm{bs}$ leads to only a 10\% decrease in co-polar coupling, demonstrating a broad optimum. 
}
   \end{figure} 

Figure~\ref{fig:mode} shows power coupling of the seven higher order modes relative to the TE$_{11}$ mode.    
The legend lists mode indices $n$, which correspond to the TE$_{11}$ mode that is anti-aligned with the probes, TM$_{01}$, 2 polarizations of TE$_{21}$, TE$_{01}$, and 2 polarizations of TM$_{11}$.  
For nearly all frequencies of interest ($f/f_\mathrm{c}<$~2.2), only two modes couple at any appreciable level.    
The TM$_{01}$ (mode 3) coupling is $>$-10~dB for $f/f_c~>$1.4, plateaus to -3.5~dB for 1.6~$<~f/f_\mathrm{c}<$~2, and reaches 0~dB at $f/f_\mathrm{c}$~=~2.26.  
The TE$_{21}$ polarization best aligned with the probes has $>-10$~dB coupling when $f/f_\mathrm{c}~>$1.7, plateaus to -6~dB for 1.8~$<~f/f_\mathrm{c}<$~2.1, and reaches 0~dB at $f/f_\mathrm{c}$~=~2.34.  
All other modes couple a negligible amount.  
To reject the TM$_{01}$ and TE$_{21}$ modes, on-chip microwave components have been implemented \cite{uyen2008magicT,mcmahon2012multi}, which exploit the even/odd mode difference between higher order modes and TE$_{11}$.    
        
\subsection{Optimal Parameter Values}
\label{sec:optimal}

In this subsection we discuss the change in metrics as a function of the most relevant geometric parameters.  
We begin with the backshort distance $z_\mathrm{bs}$, which is summarized in Fig.~\ref{fig:backshort}.
The backshort distance affects the bandwidth.  
As expected, larger backshort distances increase coupling at lower frequencies at the expense of higher frequencies and vice versa (Fig.~\ref{fig:backshort} left).  
However, as illustrated in Fig.~\ref{fig:backshort} right, a broad optimum exists.  
For example in the low band, a 25\% decrease from the optimum $z_\mathrm{bs}$ results in only a 10\% decrease in co-polar coupling.   

\begin{figure} [t]
   \begin{center}
   \begin{tabular}{c} 
   \includegraphics[width=\textwidth]{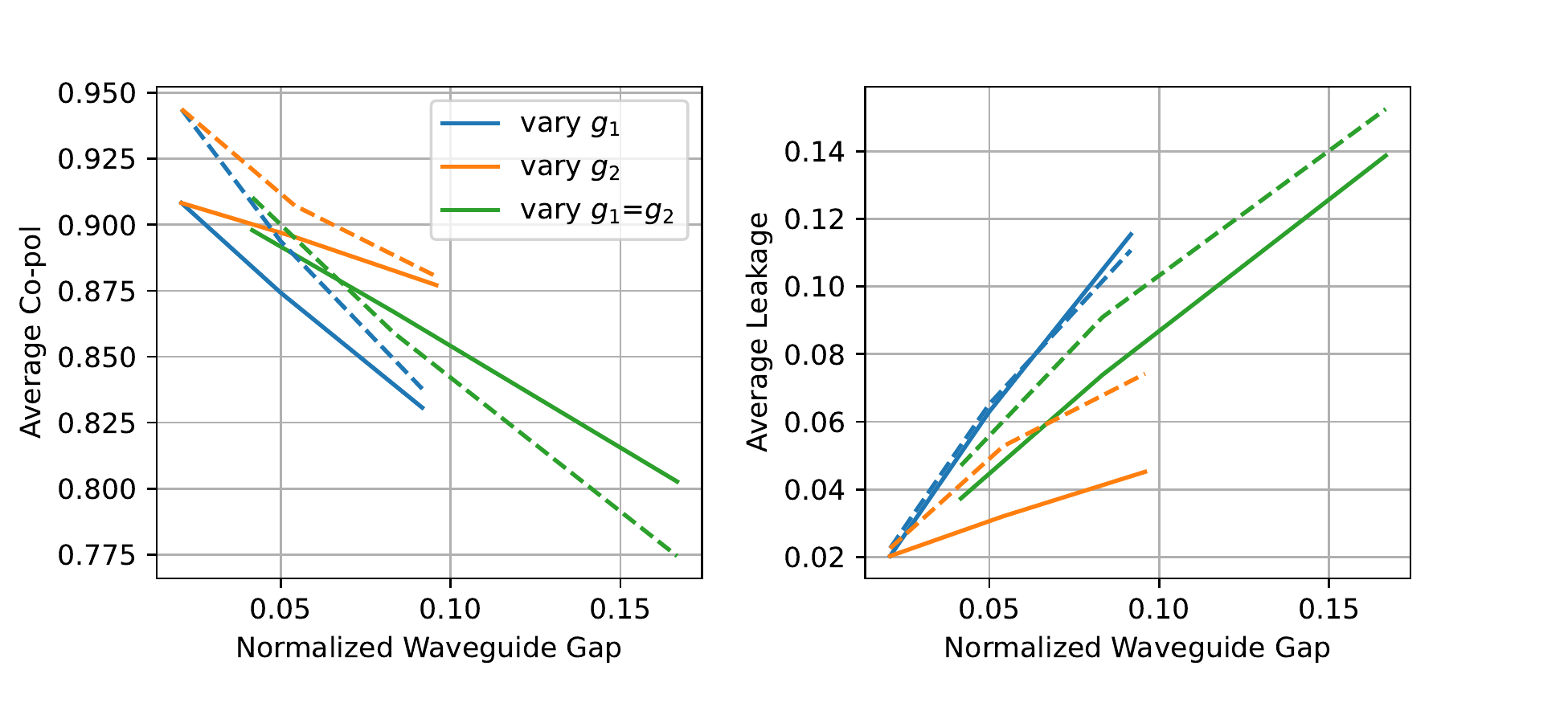}
   \end{tabular}
   \end{center}
   \caption[example] 
   { \label{fig:leakage} 
   Effect of waveguide gaps $g_{1,2}$.  
   The average co-polar coupling (left) and \leakage\ (right) is plotted as a function of normalized total waveguide gap, $(g_1+g_2)/a$.  
   The colors shown in the legend denote the different parameter sweeps.  
   The solid (dashed) lines show the average in the low (high) frequency band.
}
   \end{figure} 

As the waveguide gaps $g_{1,2}$ increase, the leakage radiation out of the waveguide increases and in turn the co-polar coupling decreases.  
We determine the co-polar and leakage radiation in three parameter sweeps.    
First, we vary the top (bottom) waveguide while the bottom (top) waveguide has fixed $g_2$=10~\micron\ ($g_1$=15~\micron).  
Modeling non-equal gaps is intentional, reflects the actual assemblies, and is the result of using wafers with standard thickness tolerance that are readily available.      
In the third case, we set $g_1$~=~$g_2$.  
We show the average \copol\ and \leakage\ within the low and high bands in Fig.~\ref{fig:leakage} as a function of normalized total waveguide gap, $(g_1+g_2)/a$.  
The \leakage\ and decrement in \copol\ are roughly linear with the waveguide gap.
We observe that leakage is more sensitive to gaps of the top waveguide rather than the backshort waveguide, particularly at lower frequencies.  
For the symmetric gap case, we see that \leakage~$\sim$~10\% when $g_1+g_2$ is 10\% of the waveguide radius.  

The membrane radius $a_c$ has no effect on the metrics to within simulation uncertainty.  
Similarly the CPW impedance has wide tolerance.  
We vary $g_\mathrm{cpw}$ from 3\micron\ to 6\micron. 
A 20\% change in $g_\mathrm{cpw}$ leads to only a 1\% change in \copol\ in both bands.
Our fiducial value produces a co-polar coupling within 1\% of the maximum in either band.    
Lastly we note that, other than affecting coupling to evanescent modes, the results are independent of the input waveguide length, as expected.  
From these parameter sweeps, we determine the optimal parameter values listed in Tab.~\ref{tab:model_parameters}. 

\subsection{Waveguide Misalignment}
\label{sec:misalignment}

\begin{figure}[t]
   \begin{center}
   \begin{tabular}{c} 
   \vspace{-0.1in}
    \includegraphics[width=0.25\textwidth]{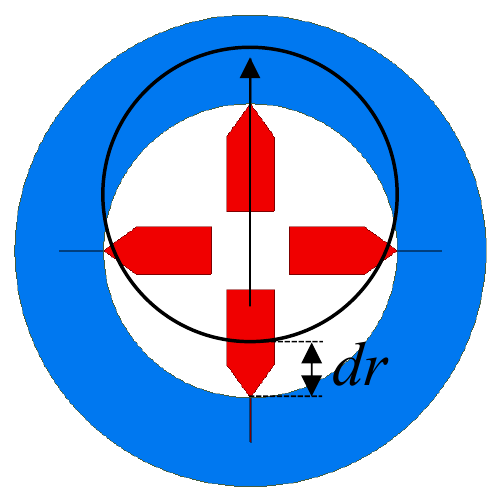}  
    \hspace{1.25in}
    \includegraphics[width=0.25\textwidth]{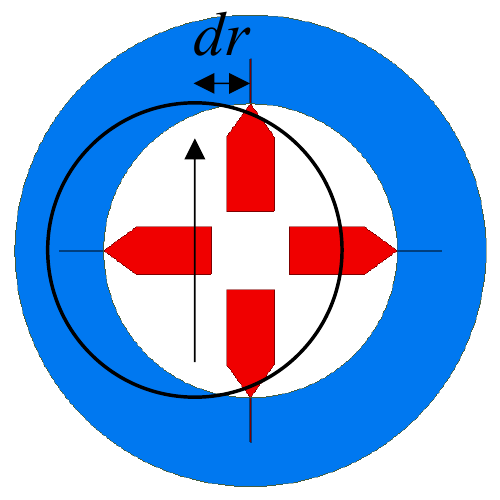}  \\
   \includegraphics[width=\textwidth]{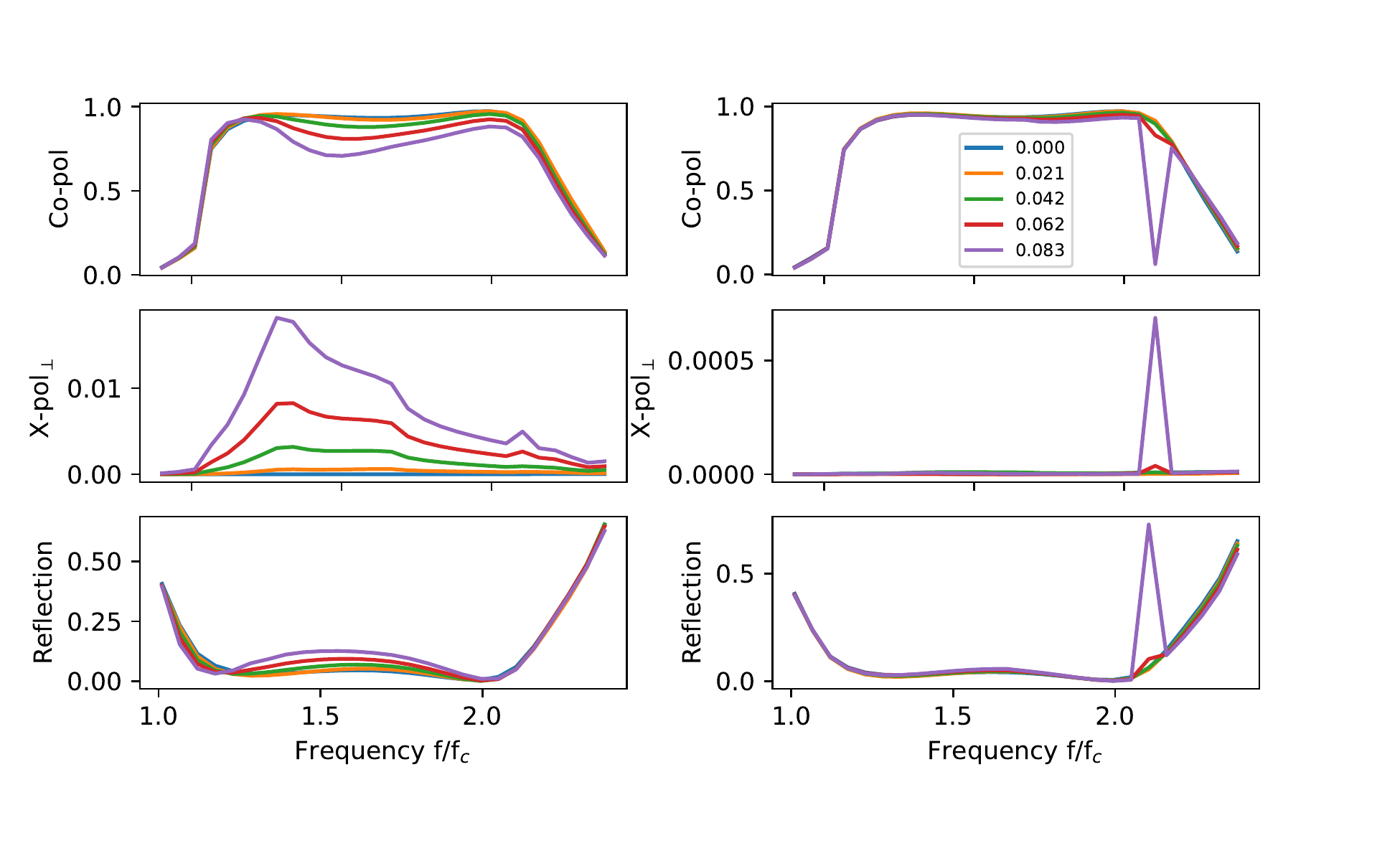}
   \end{tabular}
   \end{center}
   \vspace{-.5in}
   \caption[example] 
   { \label{fig:misalign_v_freq} 
   Effect of waveguide misalignment.  
   Results depend on the direction of misalignment with respect to the TE$_{11}$ polarization axis.  
   The planar OMT is more (less) affected by misalignment parallel (perpendicular) to the polarization axis.  
   The top row of images illustrate the misalignment geometry.  
   The left (right) column of plots show the optical coupling metrics as a function of frequency for the parallel (perpendicular) misalignment case.   
   The legend, common to all plots, lists the radial misalignment normalized to the waveguide radius $a$. 
   The notation X-pol$_\perp$ indicates that the cross-polarization belongs to the OMT probes perpendicular to the TE$_{11}$ polarization axis.  
   The results shown correspond to simulations that vary the bottom waveguide section while the top section is aligned and static.    
   The magnitude of the results vary slightly for the other parameter sweep combinations explored, but the general phenomenology holds.
}
   \end{figure}

The standard construction of the OMT uses three separate silicon parts: 
the top waveguide, the detector wafer that contains the OMT probes, and the bottom backshort waveguide.  
As such, misalignment of these parts is expected.    
In this section we determine the tolerance to radial misalignment.  
We perform parameter sweeps shifting the top and bottom waveguide sections independently, together in the same direction, and in opposite directions.  

The general effects are illustrated  Fig.~\ref{fig:misalign_v_freq}, which show that the OMT is more sensitive to misalignment in the direction parallel to the TE$_{11}$ polarization axis.  
The performance degradation is an order of magnitude less for equal radial shifts in the perpendicular direction.  
Waveguide misalignment results in decreased sensitivity, differential gain, and increased cross-polarization.  
An observed effect of lesser importance is the large reflection in a narrow band near 2.08$f/f_\mathrm{c}$, 
associated with the turn-on of TE$_{01}$ and TM$_{11}$ modes, for shifts perpendicular to the polarization axis.  

\begin{figure} [t]
   \begin{center}
   \begin{tabular}{c} 
    \includegraphics[width=\textwidth]{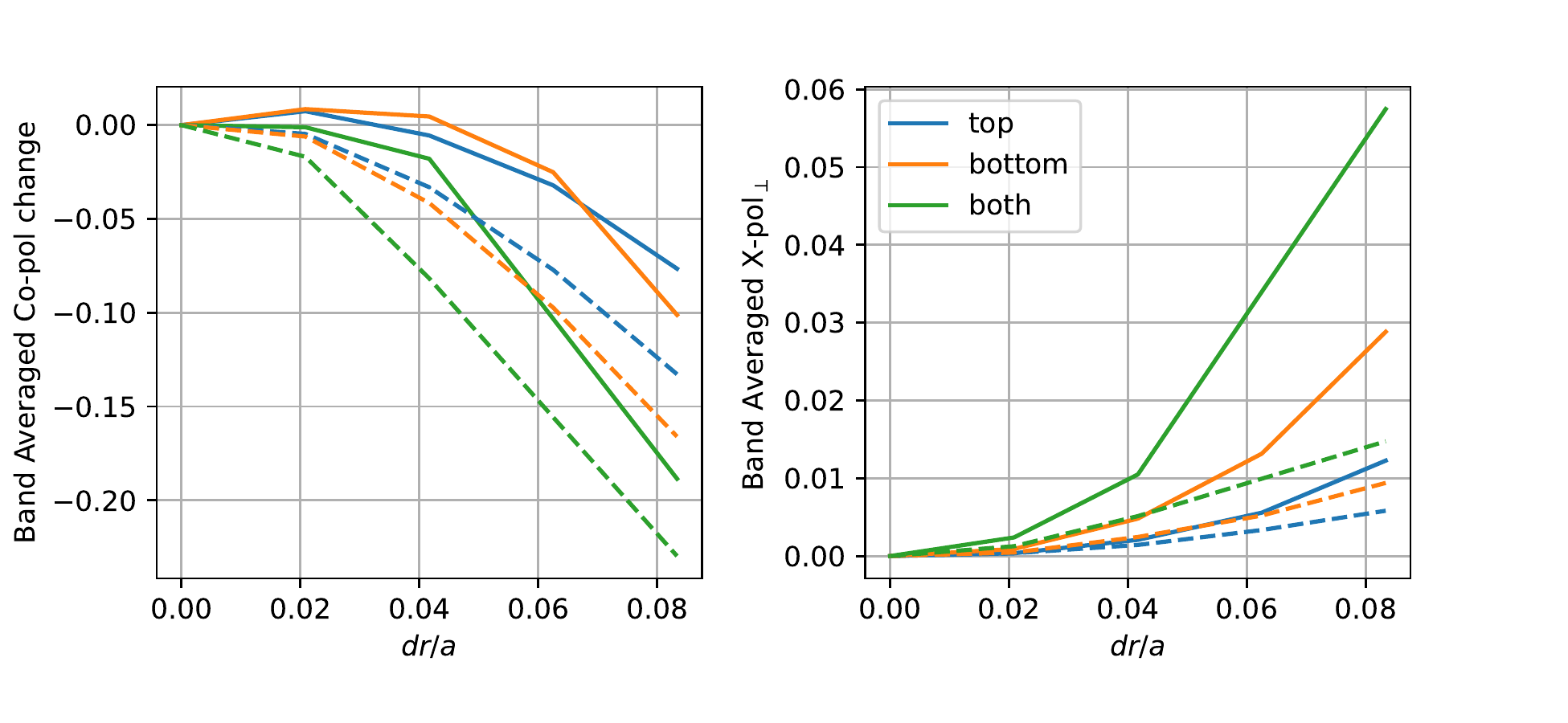}  
   \end{tabular}
   \end{center}
   \caption[example] 
   { \label{fig:misalign_band_ave} 
  Waveguide misalignment band-averaged co- and cross- polar coupling.   
  Left shows the band-averaged change in co-polar coupling from perfect alignment as a function of the normalized waveguide displacement $dr/a$ in the direction parallel to the polarization axis.  
  Right shows the band-averaged cross-polarization in probes orthogonal to the displacement direction as a function of $dr/a$.
  Colors denote the different parameter sweep variations.  
  The legend indicates the displaced waveguide part.  
  The `both' parameter sweep shifts both the top and bottom waveguide sections in the same direction.  
  The solid (dashed) line are the results for the low (high) frequency band.  
}
   \end{figure}

The band-averaged co- and cross- polar couplings are shown in Fig.~\ref{fig:misalign_band_ave}.   
Performance degradation is more sensitive to shifts of the backshort waveguide than the top waveguide when the sections are displaced independently.  
Most notably, the cross-polarization in the low frequency band is a factor of two larger.   
However, the OMT is most sensitive to shifts of both sections in the same direction.  
Shifts in opposite directions up to $dr/a$~=~0.04 fare better, and as such we do not include these results in Fig.~\ref{fig:misalign_band_ave}.

Misalignment sweeps were run with larger waveguide gaps $g_{1,2}$, and the fractional changes in the optical coupling metrics were the same as those for the nominal gap sizes to within simulation accuracy.


\subsection{`Wineglass' Probes}
\label{sec:wine}

\begin{figure} [t]
   \begin{center}
   \begin{tabular}{c} 
   \includegraphics[width=\textwidth]{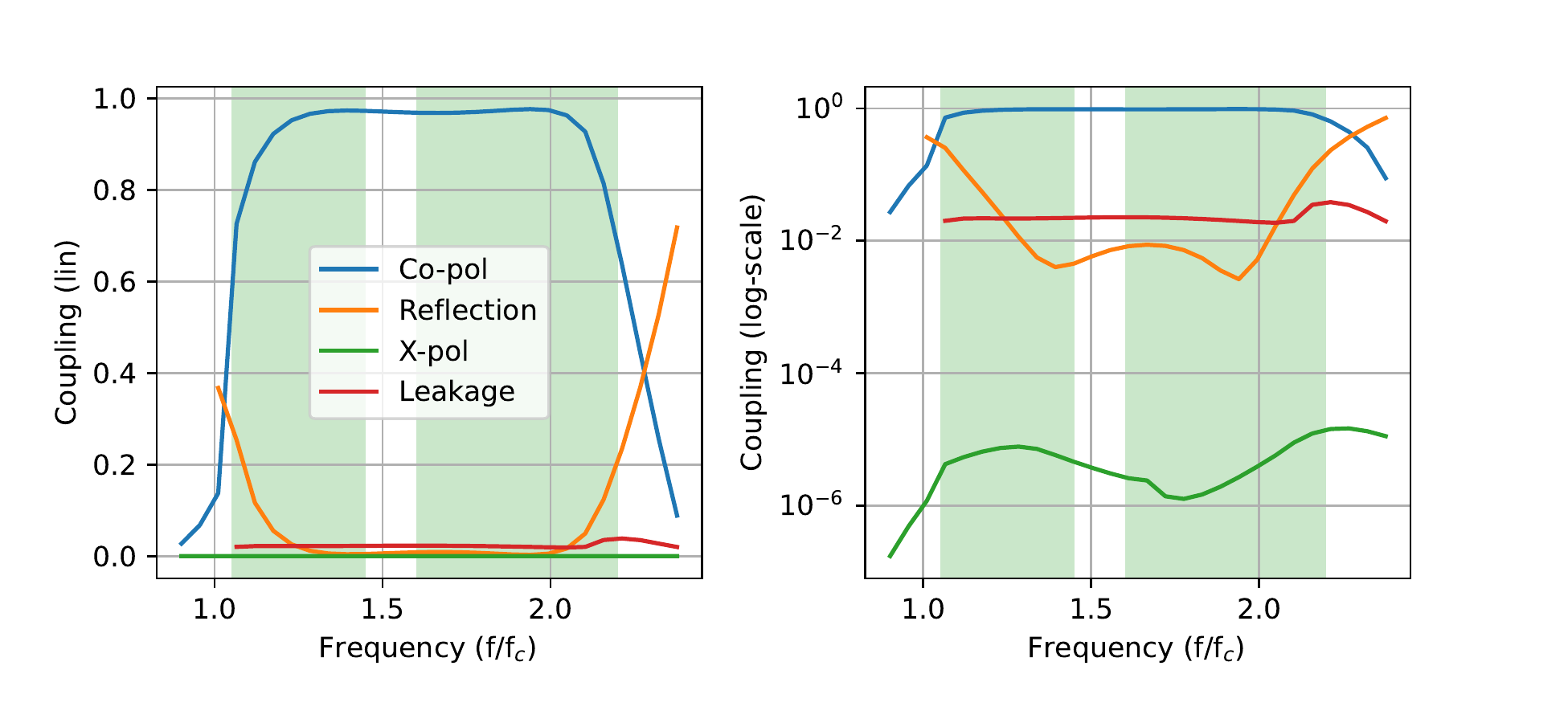}
   \end{tabular}
   \end{center}
   \caption[example] 
   { \label{fig:wine_coupling} 
Linear (left) and logarithmic (right) `wineglass' probe OMT optical coupling summary.  The bandwidth ratio is identical to the `classic'-probe OMT. }
   \end{figure}

\begin{figure}[b]
   \begin{center}
   \begin{tabular}{c} 
   \includegraphics[width=\textwidth]{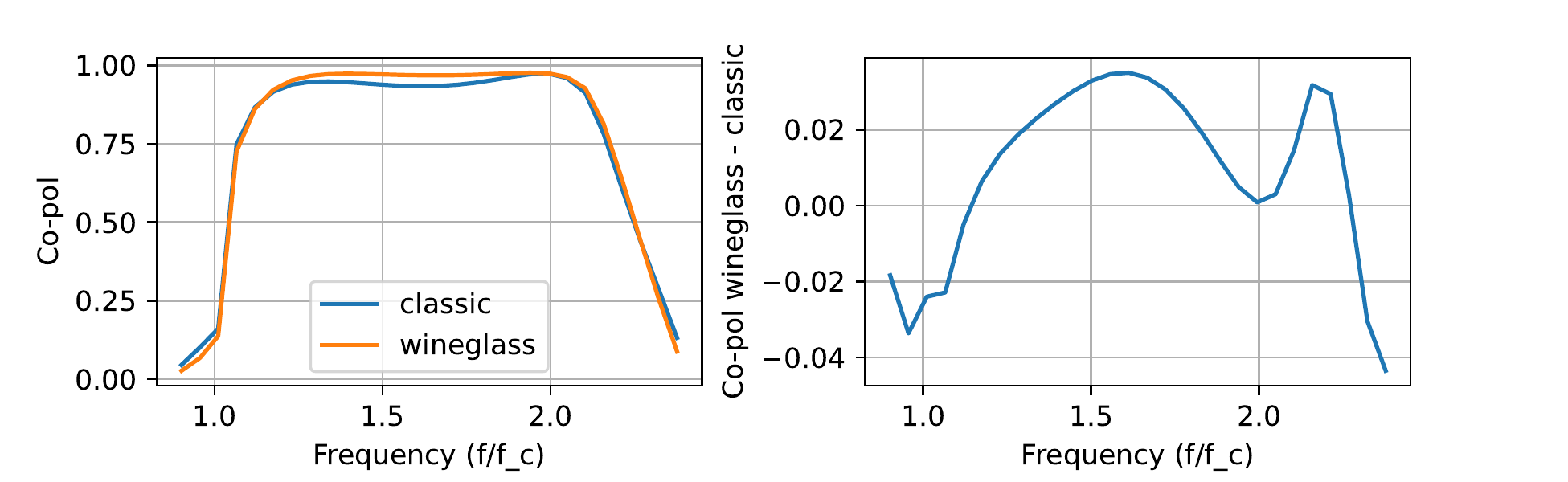}
   \end{tabular}
   \end{center}
   \caption[example] 
   { \label{fig:probe_compare} 
	Direct comparison of the classic and wineglass planar OMT.   
	The wineglass probe OMT \copol\ is on-average 2\% higher than the classic probe OMT in the range 1.2$<$f/f$_c<$2.25.}
   \end{figure}

An alternative to the classic probe geometry has been discussed in the context of CMB-S4, and we refer to this design as the `wineglass probe' due to its shape (see Fig.~\ref{fig:probe_geometry} right).  
The wineglass probe shape is parameterized by the following equation:
\begin{eqnarray}
y(x) = 0.3a(bx-1)\tanh{(-\alpha x)}.
\end{eqnarray}
Through independent parameters sweeps, we find that the values which maximize co-polar coupling are $\alpha$~=~3.6 and $b$~=~0.72.  However, a fairly broad optimum exists.   
We also find that the optimal $z_\mathrm{bs}/a$~=~0.521, slightly less than the distance in the classic probes.   
Figure~\ref{fig:wine_coupling} summarizes the wineglass probe coupling, and Fig.~\ref{fig:probe_compare} directly compares the co-polar coupling to the classic probes.  
Table~\ref{tab:ave_coupling_values} lists the band averaged metrics for the wineglass and classic probes within three frequency ranges.  

The wineglass probes provide $\sim$~2\% higher average co-polar coupling.  
The same waveguide misalignment study was carried out, and no appreciable differences were discovered relative to the results of Sec.~\ref{sec:misalignment}.

\begin{table}[t]
   \centering
   \begin{tabular}{@{} lcccccc @{}} 
      \toprule
      & \multicolumn{3}{c}{CLASSIC PROBE} &  \multicolumn{3}{c}{WINEGLASS PROBE}  \\
      
      & [1.05, 1.45] & [1.6, 2.2] & [1.2, 2.0] &  [1.05, 1.45] & [1.6, 2.2] & [1.2, 2.0] \\
      \midrule
      co-pol & 0.907 & 0.926 & 0.947 & 0.918 & 0.944 & 0.970 \\
      reflection & 0.072 & 0.045 & 0.033 & 0.058 & 0.028 & 0.008 \\ 
      leakage & 0.021 & 0.024 & 0.021 & 0.024 & 0.023 & 0.022 \\ 
            
            \bottomrule
            
   \end{tabular}
      \vspace{0.1in}
   \caption{Comparison of `classic' to `wineglass' probe band-averaged metrics for several frequency bands.  
   The frequency range is listed in the column heading as a fraction of the waveguide cut-off frequency.
  }
   \label{tab:ave_coupling_values}
\end{table}

\section{DISCUSSION}
\label{sec:discussion}

To vastly decrease the simulation volume, the excitation begins in circular waveguide.  
The model does not include a feedhorn attached to the waveguide.  
A limited set of simulations including a feedhorn profile have been carried out, and there is good agreement for the co-polar coupling.    
The main difference is that cross-polarization reaches $\sim$~-20~dB, which is expected from the horn design \cite{simon2018feedhorn}.  


On average 2\% of the power in one polarization leaks outside of the waveguide.  
While small, this power is uncontrolled and may lead to non-idealities. 
In our experience from ground-based CMB observations with planar OMTs, waveguide leakage has not been problematic.  
Nonetheless, opportunities to improve in this aspect of the design exist.
Approaches to mitigate waveguide leakage include the addition of microwave absorber, inclusion of a waveguide choke, metallic bump bonds to form a seal around the waveguide \cite{rostem2016silicon}, 
and passing the CPW probes through minimum-sized waveguide wall holes \cite{gualtieri2022optical}.  


The wineglass OMT probes provide modest gain in co-polar coupling, but we note that the reduced reflection may also help limit image `ghosting', which can be of concern for large, flat focal planes.  
Both the classic and wineglass planar OMTs have the same tolerance to waveguide misalignment.  
A good rule of thumb is to keep any radial waveguide misalignment to the probes less than 4\% of the waveguide radius.  
This criteria will ensure $<$10\% co-polar coupling reduction and that cross-polarization remains $<$-20~dB.  
The top-waveguide misalignment study presented may be a useful reference when mating the planar OMT array to metal feedhorns where differential thermal contraction is of concern.



\section{CONCLUSION}
\label{sec:conclusion}

We have presented a detailed simulation study of the optical coupling performance of octave bandwidth planar OMTs.  
Prior to this work, relevant results were limited to OMTs of 30\% fractional bandwidth.  
For the designs presented, the bandwidth ratio defined where co-polar coupling exceeds 80\% is 2.0:1.  
The average co-polar coupling, cross-polar coupling, reflection, and waveguide leakage is approximately 93\%, $<$~50~dB, 5\% and 2\%, respectively and depends slightly on the exact frequency range.  
The `wineglass' probes increase co-polar coupling and decrease reflection by $\sim$~2\% as compared to the `classic' probes.  
A good rule of thumb to avoid strong performance degradation is to align the top and bottom waveguides to the OMT probes to better than 4\% of the waveguide radius.  
We hope these results provide general guidance in the development of future focal planes for CMB measurements and beyond.

\acknowledgments 
 
Support for this work was provided in part by the NASA APRA program, proposal \#17-APRA17-0045 . 


\begin{thebibliography}{10}

\bibitem{essinger2009atacama}
Essinger-Hileman, T., et~al. ``The Atacama B-mode Search: CMB polarimetry with transition-edge-sensor bolometers,'' AIP {\it conference proceedings}  {\bf 1185}(1),  494--497, (2009).

\bibitem{stanchfield2016development}
Stanchfield, S., et~al., ``Development of a microwave SQUID-multiplexed TES array for MUSTANG-2,'' {\em JLTP}~{\bf 184}(1), 460--465 (2016).

\bibitem{austermann2012sptpol}
{Austermann}, J.~E., et~al., ``{SPTpol:
  an instrument for CMB polarization measurements with the South Pole
  Telescope},''  {\em SPIE proceedings} {\bf 8452} (Sept. 2012).

\bibitem{niemack2010actpol}
{Niemack}, M.~D., et~al., ``{ACTPol: a polarization-sensitive receiver for the Atacama Cosmology
  Telescope},''  {\em SPIE proceedings} {\bf 7741} (July 2010).

\bibitem{hubmayr2016spider}
Hubmayr, J., et~al., ``Design of 280 GHz feedhorn-coupled TES arrays for the balloon-borne
  polarimeter SPIDER," {\em SPIE proceedings} {\bf 9914},  99140V, (2016).

\bibitem{essinger2014class}
Essinger-Hileman,T., et~al., ``CLASS: the
  cosmology large angular scale surveyor,'' {\em SPIE proceedings}  {\bf 9153},  91531I, (2014).

\bibitem{mcmahon2009planar}
McMahon, J., {\em AIP Conference
  Proceedings},   {\bf 1185}(1),  490--493,
  American Institute of Physics (2009).

\bibitem{grimes2007compact}
Grimes, P., King, O., Yassin, G., and Jones, M., ``Compact broadband planar
  orthomode transducer,'' {\em Electronics Letters}~{\bf 43}(21),  1146--1148
  (2007).

\bibitem{mcmahon2012multi}
McMahon, J., et~al., ``Multi-chroic feed-horn
  coupled TES polarimeters,'' {\em Journal of Low Temperature Physics}~{\bf
  167}(5-6),  879--884 (2012).

\bibitem{thornton2016atacama}
Thornton, R., et~al., ``The Atacama Cosmology Telescope:
  the polarization-sensitive ACTpol instrument,'' {\em The Astrophysical
  Journal Supplement Series}~{\bf 227}(2),  21 (2016).

\bibitem{henderson2016advanced}
Henderson, S., et~al., ``Advanced
  ACTPol cryogenic detector arrays and readout,'' {\em JLTP}~{\bf 184}(3),  772--779 (2016).

\bibitem{ade2019simons}
Ade, P., et~al., ``The Simons
  Observatory: science goals and forecasts,'' {\em JCAP}~{\bf 2019}(02),  056 (2019).

\bibitem{li2019probing}
Li, H., et~al., ``Probing primordial gravitational waves:
  Ali cmb polarization telescope,'' {\em National Science Review}~{\bf 6}(1),
  145--154 (2019).

\bibitem{abazajian2019cmb}
Abazajian, K., et~al., ``CMB-S4
  science case, reference design, and project plan,'' {\em arXiv preprint
  arXiv:1907.04473}  (2019).

\bibitem{mazumi2020litebird}
Hazumi, M., et~al.,
  ``{LiteBIRD satellite: JAXA's new strategic L-class mission for all-sky
  surveys of cosmic microwave background polarization},'' {\em SPIE proceedings} {\bf 11443},  431 -- 450, (2020).

\bibitem{ward2016mechanical}
Ward, J.~T., et~al.,
  ``Mechanical designs and development of tes bolometer detector arrays for the
  advanced actpol experiment,'' {\em SPIE proceedings} {\bf 9914},  794--802, (2016).

\bibitem{duff2016advanced}
Duff, S.~M., et~al., ``Advanced ACTPol
  multichroic polarimeter array fabrication process for 150 mm wafers,'' {\em
  JLTP}~{\bf 184}(3),  634--641 (2016).

\bibitem{uyen2008magicT}
U-yen, K., Wollack, E.~J., Papapolymerou, J., and Laskar, J., ``A broadband
  planar magic-t using microstrip slotline transitions,'' {\em IEEE
  Transactions on Microwave Theory and Techniques}~{\bf 56}(1),  172--177
  (2008).

\bibitem{simon2018feedhorn}
Simon, S.~M., et~al., ``Feedhorn
  development and scalability for Simons Observatory and beyond,'' {\em SPIE proceedings} {\bf 10708},  787--798, (2018).

\bibitem{rostem2016silicon}
Rostem, K., et~al., ``{Silicon-based antenna-coupled
  polarization-sensitive millimeter-wave bolometer arrays for cosmic microwave
  background instruments},''  {\em SPIE proceedings} {\bf 9914},  54 -- 63, (2016).

\bibitem{gualtieri2022optical}
Gualtieri, R., Barry, P.~S., Cecil, T., Bender, A.~N., Chang, C.~L., Hood,
  J.~C., Lisovenko, M., and Yefremenko, V.~G., ``Optical leakage mitigation in
  ortho-mode transducer detectors for microwave applications,'' {\em JLTP}, 1--8 (2022).

\end{thebibliography}

\end{document}